\documentclass{article}
\usepackage{spconf,amsmath,graphicx}
\usepackage{multirow}
\usepackage{booktabs}
\usepackage{xcolor}
\usepackage[colorlinks]{hyperref}
\usepackage{subcaption}

\usepackage[vlined,linesnumbered,ruled]{algorithm2e}
\usepackage[noadjust]{cite}

\def\d{{\mathrm d}}

\title{uSee: Unified Speech Enhancement and Editing with Conditional Diffusion Models}
\name{\begin{tabular}{c}Muqiao Yang$^{1\dagger}$, Chunlei Zhang$^{2}$, Yong Xu$^{2}$, Zhongweiyang Xu$^{3}$, Heming Wang$^{4}$, Bhiksha Raj$^{1,5}$, Dong Yu$^{2}$ \end{tabular}\thanks{$^\dagger$Work performed during internship at Tencent AI Lab.}}
\address{   
$^1$ Carnegie Mellon University, $^2$ Tencent AI Lab \\
$^3$ University of Illinois Urbana-Champaign, $^4$ The Ohio State University, $^5$ MBZUAI}
\begin{document}
\ninept
\maketitle
\begin{abstract}
Speech enhancement aims to improve the quality of speech signals in terms of quality and intelligibility, and speech editing refers to the process of editing the speech according to specific user needs. In this paper, we propose a Unified Speech Enhancement and Editing (uSee) model with conditional diffusion models to handle various tasks at the same time in a generative manner. Specifically, by providing multiple types of conditions including self-supervised learning embeddings and proper text prompts to the score-based diffusion model, we can enable controllable generation of the unified speech enhancement and editing model to perform corresponding actions on the source speech. Our experiments show that our proposed uSee model can achieve superior performance in both speech denoising and dereverberation compared to other related generative speech enhancement models, and can perform speech editing given desired environmental sound text description, signal-to-noise ratios (SNR), and room impulse responses (RIR). Demos of the generated speech are available at \url{https://muqiaoy.github.io/usee}.

\end{abstract}
\begin{keywords}
Speech Enhancement, Speech Editing, Diffusion Models, Generative Models
\end{keywords}

\newcommand{\ra}[1]{\renewcommand{\arraystretch}{#1}}

\makeatletter
\newcommand{\removelatexerror}{\let\@latex@error\@gobble}
\makeatother

\section{Introduction}
\label{sec:intro}

Speech enhancement is a critical front-end component in speech processing, aiming to recover the clean speech signals that are potentially corrupted by acoustic noises or reverberation \cite{loizou2007speech}. Many real-world applications, including automatic speech recognition (ASR), speaker verification (SV), and telecommunication systems \cite{shi2021improving,shi2022investigation,zhang2021towards}, rely on the accuracy and reliability of the captured speech signal. In this context, speech enhancement plays a pivotal role in developing advanced algorithms and techniques to enhance the quality and intelligibility of speech signals, making them more suitable for subsequent processing and analysis. 

Deep Neural Network (DNN) based speech enhancement methods have been widely investigated with mapping-based and masking-based methods \cite{hu2020dccrn, defossez2020real, luo2019conv}. Mapping-based methods provide targets as the time-frequency (T-F) domain representation of the clean speech, so that the DNN learns to predict the mapping from noisy to clean signal as a regression problem \cite{lu2013speech, xu2014regression}. On the other hand, masking-based methods build the training targets of the DNN as a binary T-F mask to decide whether each portion is noise or speech, thus making it a classification problem \cite{narayanan2013ideal}. However, there still exists room for improvement in such techniques, especially when addressing real-world recording scenarios. One major challenge is that multiple types of distortions may happen in the real world, such as background noise, reverberation, and interference, etc., all of which can degrade the quality of the acquired audio. Generation-based speech enhancement models have gathered increasing attention in the community, given their capability to generate realistic speech by matching its overall distribution to clean speech instead of optimizing the DNN with a point-wise objective function \cite{pascual2017segan, lu2022conditional}. However, most of them are still designed to solve one single enhancement task, and may not generalize to multiple types of distortions in real applications \cite{wang2023speechx}.


Besides speech enhancement, the demand for speech editing has surged across various applications, including short video editing, immersive media experiences, etc \cite{wang2023audit,wang2023speechx}. Speech editing refers to the process of performing editing operations on speech according to user instructions, e.g., adding background sound or reverberation effects to the source speech. This motivates the emergence of some generation-based speech editing systems, which leverage text transcriptions to edit speech according to specific user needs \cite{tan2021editspeech, tae2021editts}. Nevertheless, most existing approaches focus on only one specific editing task, restricting the model from generating edited speech with flexible human instructions. Wang et al. \cite{wang2023audit} proposed an audio editing tool for multiple tasks, but they are mainly focused on audio tasks instead of speech editing, so they did not provide fine-grained control like signal-to-noise ratios (SNR) or room sizes via text instructions. Meanwhile, some text-to-audio generation methods are proposed in \cite{liu2023audioldm, kreuk2022audiogen}.

In this work, we propose a Unified Speech Enhancement and Editing (uSee) model that can perform speech denoising, dereverberation and controlled speech editing, given source speech and companion text prompt instructions. The backbone of the proposed model is based on stochastic differential equations (SDE) for score-based conditional diffusion models \cite{song2020score}. By treating both speech enhancement and speech editing as generation tasks, our uSee model enables controllable speech generation conditioned on specific acoustic and textual prompts. In summary, our uSee model makes contributions as follows. First, we investigate the effect of the linear and exponential interpolation between source and target spectrograms as a condition of the diffusion model. Moreover, the proposed model can manipulate its generation behavior by acoustic prompts containing self-supervised learning (SSL) embeddings from the source speech, and textual prompts describing the semantic information of requirements desired by users. By combining multiple conditions as the conditional input, our uSee model presents the capability of controllable speech generation for both speech enhancement and speech editing tasks. 
Our experiments demonstrate that the proposed uSee model can not only effectively generate enhanced speech by removing background noise and reverberation, but also generate edited speech with specific background sound types, signal-to-noise ratios (SNR), and room impulse responses (RIR) described by a textual condition.
\begin{figure*}
    \centering
    \includegraphics[scale=0.6]{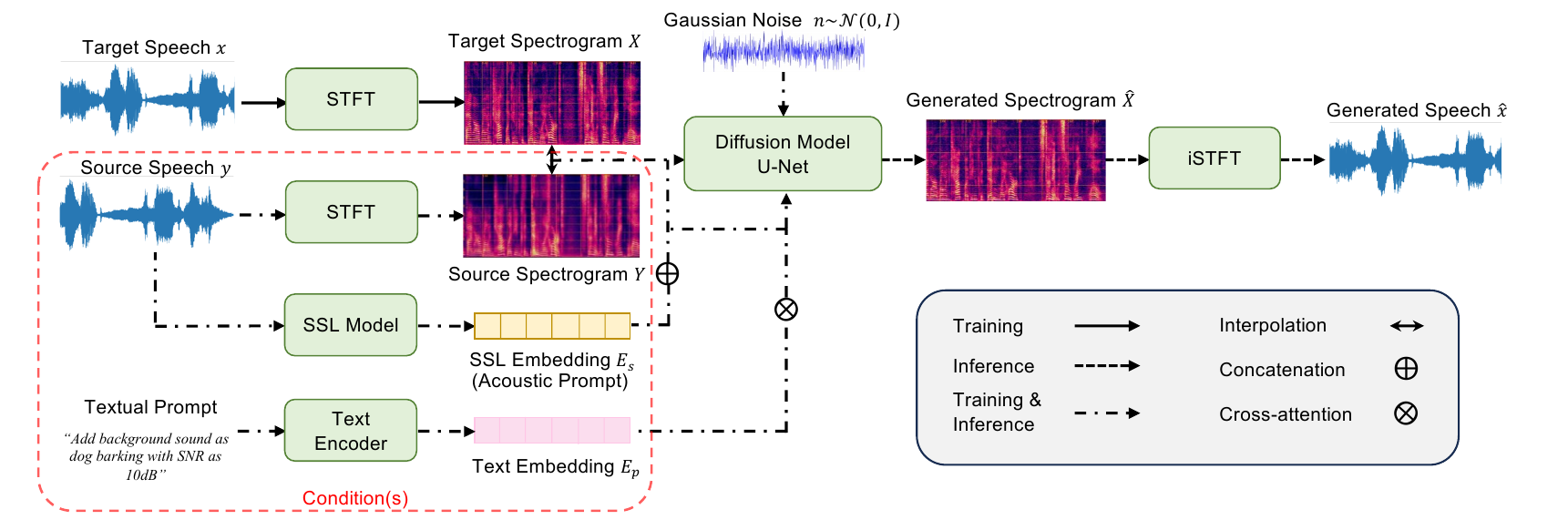}
        \vspace{-3mm}
    \caption{Overview of the proposed unified speech enhancement and editing (uSee) model.}
    \label{fig:teaser}
    \vspace{-5mm}
\end{figure*}



\section{Related Works}
\label{sec:related_work}

Among generation-based speech enhancement models, diffusion models have been studied in an increasing number of works \cite{ho2020denoising, song2020denoising}. The concept of the diffusion process is to continuously add random noise perturbations to data from a prior distribution, and then a network is trained to learn to reverse the diffusion process to recover desired data samples from the noise. Lu et al. \cite{lu2022conditional} used a conditional diffusion probabilistic model to synthesize enhanced speech in the time domain. Richter et al. \cite{richter2023speech} proposed an SDE-based diffusion process in the complex T-F domain, where a score matching objective is applied to estimate the score of the perturbed data distribution. In speech editing, Wang et al. \cite{wang2023audit} presented an audio edit tool with a diffusion model to generate edited speech according to human instructions. However, existing studies mostly focus on only one or limited aspects of tasks. There has been no prior research on a unified model to handle various speech enhancement and editing tasks at the same time, where both acoustic and textual information are provided as conditions to enable fine-grained control of the generation process.

\vspace{-1mm}
\section{Method}
\label{sec:method}

In this section, we will introduce how conditional diffusion models work with speech enhancement and editing in our proposed uSee model. We will first begin by illustrating the overall architecture of our proposed method and then describe the score-based conditional diffusion process. Then we summarize how controllable generations can be achieved with different conditions in our uSee model.

\vspace{-2mm}
\subsection{Pipeline Overview}
\label{subsec:pipeline}

The process of speech enhancement and editing can be regarded as conditional generation tasks in nature, where the corrupted speech in speech enhancement or the clean speech in speech editing is provided with the generation model as conditions. For the purpose of simplicity, we will name such raw speech segments in input conditions as source speech, and the resulting speech that the user desires as target speech. Note that target speech is only accessible during training. We denote target speech as $x$ and source speech as $y$.

We illustrate the overview of the architecture of our proposed uSee model in Figure \ref{fig:teaser}. The network architecture of the diffusion model is based on a multi-resolution U-Net structure \cite{ronneberger2015u}. For data representation, we first employ short-time Fourier Transform (STFT) to convert source speech $y$ and target speech $x$ into T-F domain representations, and use the resulting spectrogram $X$ and $Y$ as the input and conditions to the diffusion model. Note that the target speech $x$ is only available during the training process, denoted by solid arrows. During the forward process at each time step, a Gaussian white noise will be continuously added to the target spectrogram $X$. As the output of the conditional diffusion model during inference, an estimated spectrogram $\hat{X}$ will be iteratively generated through the reverse process, such that the raw target spectrogram is expected to be recovered. As the final step, an inverse STFT (iSTFT) is further applied to $\hat{X}$ to convert it back to the generated speech $\hat{x}$. This step only happens during the inference stage, denoted by dashed arrows.

\vspace{-2mm}
\subsection{Controllable Speech Generation via Conditional Inputs}


Controlling the behavior of generative models is important to make them applicable in real-world applications. To demonstrate the necessity of all conditions to achieve high-quality controllable generation, we further investigate the effect of different combinations of conditions in our uSee model, as shown in the block of Condition(s) in Figure \ref{fig:teaser}. The integration of multiple conditions plays a pivotal role in achieving fine-grained control over the generation process, allowing us to manipulate various aspects of the generated speech, such as noise reduction, reverberation removal, and the introduction of specific background sounds and RIRs.

\noindent \textbf{Source Spectrograms.} As we have mentioned in Section \ref{subsec:pipeline}, both speech enhancement and editing tasks can be considered as conditional generation of target speech given a source speech input. Source spectrograms are a critical component in the conditional controls, as they provide the basic information for the model to generate the target speech. Since the reverse process of a diffusion model is to recover clean speech from a Gaussian noise with iterative steps, the condition that each step depends on should not be identical. An interpolation between source and target spectrograms with changing weights across steps can be used as the condition, where the weight of source spectrograms is higher in earlier steps, and becomes gradually lower in later steps. We have also investigated different interpolations such as linear and exponential interpolation, which we will show in Section \ref{sec:exp}.

\noindent \textbf{Acoustic Prompts.} We apply an SSL embedding $E_{s}$ from the source speech as one of the conditions to the diffusion model. The SSL embedding is expected to provide further guidance to achieve controllable generation during the condition-dependent trajectory from $y$ to $x_0$. Here, we employ the STFT with the same configuration as the HuBERT \cite{hsu2021hubert} pretraining, where the hop length is $320$. Therefore, the mixture of spectrogram and SSL embeddings from the same source-target speech segment pairs will consist of the same number of frames within one fixed window length, and we concatenate them in a frame-by-frame correspondence.  

\begin{figure}[t]
    \centering
    \includegraphics[scale=0.43]{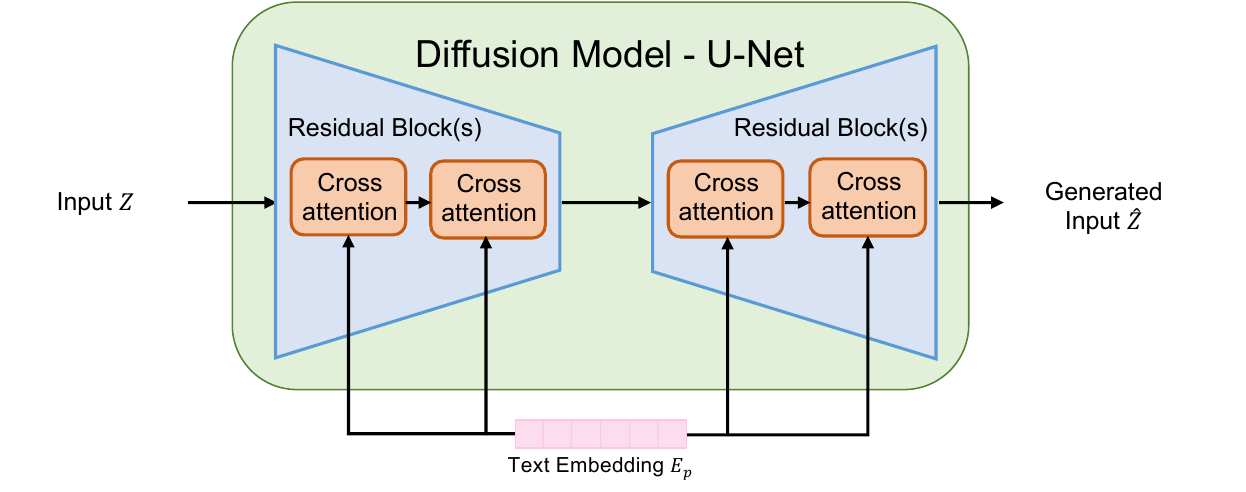}
     \vspace{-3mm}
    \caption{Illustration of cross-attention layers for textual prompts as conditions in the diffusion model.}
    \vspace{-5mm}
    \label{fig:attn}
\end{figure}

\noindent \textbf{Textual Prompts.} In addition to the acoustic prompts from the SSL embeddings, we further introduce textual prompts to the diffusion model to enable more fine-grained control of the generation process. For example, in speech enhancement, we may want to use textual prompts to inform the model whether to perform speech denoising or speech dereverberation to the source speech, or even both. Moreover, in speech editing, we will want to control the specific background sound type, SNR and RIR to be simulated, and the detailed instruction can be included in the format of textual prompts. We employ a text encoder to extract the text embeddings $E_{p}$ from the textual prompts. The text embeddings will be utilized multiple times inside the U-Net structure, where the diffusion model applies cross-attention mechanism to use the text information to steer the process to generate the audio containing the specific information in the textual prompts. A more detailed illustration is shown in Figure \ref{fig:attn}. At the output level of each residual block in the U-Net, a cross-attention layer will be applied between the text embedding $E_p$ and the input $Z$. Note that $Z$ has already contained the information from acoustic prompts before the cross-attention layers. By doing so, the generated spectrogram $\hat{Z}$ is expected to accommodate semantic information contained in the text prompts, thus achieving the fine-grained controllable generation.

\vspace{-3mm}
\subsection{Score-based Conditional Diffusion}
\label{subsec:score_cd}
\vspace{-1mm}

We build the backbone of our model based on score-based diffusion models, where the idea is to use stochastic differential equations to estimate the score of the perturbed data in generative modeling. In our uSee model, we incorporate a combination of conditions into the forward process and reverse process in the score-based diffusion model, such that the model can be enabled with controllable generation to generate enhanced or edited speech.

\vspace{-3mm}
\subsubsection{Forward Process}  
\vspace{-1mm}

During training, the forward process of score-based diffusion is to progressively add increasing levels of Gaussian noise to the target speech $x$ in the form of SDE. Given a diffusion process with $T$ steps, following \cite{song2020score}, the forward process can be defined as $\{x_t\}_{t=0}^T$ that can be modeled in the form of SDE:

\begin{equation}
    \d x_t = f(x_t, t)\d t + g(t)\d \omega
\label{eq:1}
\end{equation}

\noindent where $f$ and $g$ are the drift and diffusion coefficients, and $\omega$ denotes the standard Wiener process. Different from the interpolation between noisy speech $y$ and clean speech $x$ in \cite{lu2022conditional}, here the drift $f(x_t, t) = \gamma(y-x_t)$ where $\gamma$ is to control the level of transition between $x$ and $y$. The diffusion coefficient $g(t) = \sigma_{\mathrm{min}} (\frac{\sigma_{\mathrm{max}}}{\sigma_{\mathrm{min}}})^t \sqrt{2\log (\frac{\sigma_{\mathrm{max}}}{\sigma_{\mathrm{min}}})}$  is to control the variance schedule of the Gaussian white noise to be added, and $\sigma_{\mathrm{max}}$ and $\sigma_{\mathrm{min}}$ are both hyperparameters. Overall, the SDE defines the Gaussian process $\{x_t\}_{t=0}^T$. By default, the distribution from which each state is sampled can be defined as 

    \vspace{-4mm}
\begin{equation}
\begin{aligned}
    & x_0\sim P_0(x), \\
    & x_t\sim P_{0t}(x_t | x_0, y) = \mathcal{N}(x_t;\mu(x_0, y, t), \sigma(t)^2 \mathbf{I})
\end{aligned}
\label{eq:2}
\end{equation}


\noindent where $P_0$ is the distribution for raw target speech, and $P_{t}$ is called the perturbation kernel given an arbitrary time step $t \in [0, T]$. $I$ denotes the identity matrix. According to \cite{sarkka2019applied}, the closed form solution for the mean $\mu(x_0, y, t)$ and variance $\sigma(t)^2$ can be derived as:

    \vspace{-2mm}
\begin{equation}
\begin{aligned}
    & \mu(x_0, y, t) = e^{-\gamma t}x_0 + (1 - e^{-\gamma t})y \\
    & \sigma(t)^2 = \frac{\sigma_{\mathrm{min}}^2 ((\frac{\sigma_{\mathrm{max}}}{\sigma_{\mathrm{min}}})^{2t} - e^{-2\gamma t}) \log (\frac{\sigma_{\mathrm{max}}}{\sigma_{\mathrm{min}}})}{\gamma + \log (\frac{\sigma_{\mathrm{max}}}{\sigma_{\mathrm{min}}})}
\end{aligned}
\label{eq:3}
\end{equation}

By applying the drift coefficient $f$ in Equation \ref{eq:2}, the mean of the distribution at each step $x_t$ achieves the exponential decay from $x_0$ to $y$ in the forward process as shown in Equation \ref{eq:3}. At each time step $t$, $x_t=\mu(x_0, y, t) + z \cdot \sigma(t)^2$ will be sampled based on the marginal probability density function and then used as the conditional input to the diffusion model, where $z\in\mathcal{N}(0, \mathbf{I})$. We will further demonstrate the performance comparison between the linear interpolation and the exponential interpolation in Section \ref{sec:exp}.

\vspace{-1mm}
\subsubsection{Reverse Process}  
\vspace{-1mm}

The objective of the reverse process of diffusion is to recover the corrupted target speech $x_t$ back to the raw target speech $x_0$. Specifically, in score-based conditional diffusion, a score model $s_{\boldsymbol{\theta}}$ is trained to estimate $\nabla \log P_{0t}(x_t | x_0, y, c)$, the gradient of the log probability density of the perturbation kernel at step $t$, such that the reverse process can be modeled from $P_{0t}$ to $P_0$. Here $c$ refers to the additional condition that the model is conditioned on at time step $t$. According to \cite{anderson1982reverse}, the SDE in the forward process defined in Equation \ref{eq:1} has the reverse-time SDE as:

    \vspace{-2mm}
\begin{equation}
\begin{aligned}
    \d x_t &= [f(x_t, t) - g(t)^2 \nabla \log P_{0t}(x_t | x_0, y, c)]\d t + g(t) \d \hat{\omega} \\
    &= [f(x_t, t) - g(t)^2  s_{\boldsymbol{\theta}}(x_t, y, t, c)]\d t + g(t) \d \hat{\omega}
\end{aligned}
\end{equation}

\noindent which needs to be solved by numerical SDE solvers like annealed Langevin Dynamics \cite{song2019generative}. For the condition $c$, here we employ self-supervised learning (SSL) embeddings $E_{s}$ extracted from HuBERT \cite{hsu2021hubert} on the source speech, such that it provides additional guidance to the generation process in terms of phonetic and acoustic information. 

During the inference process, since the model has no access to the ground-truth target speech $x$, we first begin the initial state $x_T$ by sampling as $x_T\sim \mathcal{N}(x_T; y, \sigma(T)^2 \mathbf{I})$. Then the reverse process can be iteratively solved from $t=T$ to $t=0$. Similar to the forward process, across time steps, the mean of the distribution also experiences an exponential decay from $y$ to $\hat{x}$, where $\hat{x}$ is the generated estimate of the source speech.

\vspace{-3mm}
\section{Experiments}
\label{sec:exp}

\begin{table}[t]

    \centering
         \resizebox{8.5cm}{!}{
    \begin{tabular}{l|c|c|c|c}
    \toprule
    Model & WB-PESQ & NB-PESQ & STOI & DNSMOS \cite{reddy2021dnsmos} \\
    \midrule
    \midrule
        Noisy & 1.46 & 2.05 & 0.61 & 3.26 \\
        cDiffuSE \cite{lu2022conditional} & 1.75 & 2.47 & 0.69 & 3.48  \\
        SGMSE \cite{richter2023speech} & 1.91  & 2.66 & 0.72 & 3.65  \\
        \midrule
        uSee (ours) & & & & \\
        \ \  linear interp. & 1.76 & 2.49 & 0.67 & 3.50 \\
        \ \  exponential interp. & 2.01 & 2.72 & 0.71 & 3.68 \\
        \ \  \ + acoustic prompts & \multirow{1}{*}{2.15} & \multirow{1}{*}{2.86} & \multirow{1}{*}{0.77} & \multirow{1}{*}{3.80} \\
        \midrule
        \ \  \ \ + textual prompts & \multirow{1}{*}{\textbf{2.20}} & \multirow{1}{*}{\textbf{2.88}} & \multirow{1}{*}{\textbf{0.80}} & \multirow{1}{*}{\textbf{3.86}} \\
    \bottomrule
    \end{tabular}}
     \vspace{-3mm}
    \caption{Quantitative evaluation results on joint speech denoising and dereverberation of our uSee model with different conditions.}
    \vspace{-4mm}
    \label{tab:enhan_res}
\end{table}

\subsection{Datasets and Simulation}
\vspace{-2mm}

Since the objective of our uSee model is unified and multi-fold, we would like to enable infinite variants of our source and target speech by simulating them with different sets of clean speech, environmental noises, and RIRs. We choose the clean speech set as LibriTTS \cite{zen2019libritts}, noise set as the balanced training set from AudioSet \cite{gemmeke2017audio}, and RIR dataset as the Room Impulse Response and Noise Database \cite{ko2017study}. Clean speech is resampled to 16kHz before simulation. For speech enhancement tasks, we first consider denoising and deverberation as separate subtasks. We simulate noisy audios with clean speech and a random noise type via a specific SNR, and reverberant audios with clean speech and a randomly sampled RIR from large, medium, and small rooms. For speech editing, we apply a similar simulation process, while the source speech is the clean speech and the target speech is referred to as the mixture of clean speech and specific types of background sound or reverberations.

\vspace{-2mm}
\subsection{Implementation Details}

During speech simulation, we add to the clean speech with either a background sound or a room impulse response. If the clean speech is shorter than the background sound, instead of clipping the sound, we will randomly insert the clean speech into one interval of the background sound. The motivation is that we will utilize text information to generate speech with specific sound types, and in this way, the information representing the sound type will be fully preserved. If the clean speech is longer, the background sound will be repeated until it reaches the same length as the clean speech. 

The additive background sound is mixed with SNR ranging from $[0, 15]$dB, and the RIR is selected from rooms with different volumes including large, medium, and small sizes. We associate each pair of simulated source and target speech with a textual prompt. The textual prompt describes whether the objective is to remove noise, remove reverberation, add background sound, or add room impulse responses to the source speech. For example, as shown in Figure \ref{fig:teaser}, one prompt could be \textit{``Add background sound as dog barking with SNR as 10dB''}. This indicates that the objective of this command is to add dog barking as background sound to the source speech, such that the SNR is desired to be $10$dB.

For the conditions, we choose the SSL model as HuBERT \cite{hsu2021hubert} and the text encoder as T5 \cite{raffel2020exploring}. We extract the output from the last layer of HuBERT as our SSL embeddings, which will be the acoustic prompts. We will evaluate and compare the performance of different models with conditions used in our uSee model in Section \ref{subsec:exp_res}.

\vspace{-2mm}
\subsection{Experimental Results and Analysis}
\label{subsec:exp_res}

To demonstrate the effectiveness of our proposed uSee model on various tasks, we will show the evaluation results of speech enhancement and editing separately. For speech enhancement, the quantitative evaluation results are shown in Table \ref{tab:enhan_res}. We report evaluation metrics including both intrusive metrics such as PESQ and STOI, and non-intrusive metric DNSMOS \cite{reddy2021dnsmos}. Notably, our uSee model demonstrates superior performance when compared to other diffusion-based generative speech enhancement models, such as cDiffuSE \cite{lu2022conditional} and SGMSE \cite{richter2023speech}, across all evaluated metrics. Furthermore, we conduct ablation experiments to investigate the impact of different conditions to the diffusion model. The results show that both acoustic and textual prompts have an effect on boosting the speech enhancement performance of uSee model individually. Consequently, the optimal performance is achieved through exponential interpolation, utilizing SSL embeddings as the acoustic prompt and text embeddings as the textual prompt.

\begin{figure}[t]
    \centering
    \includegraphics[scale=0.39]{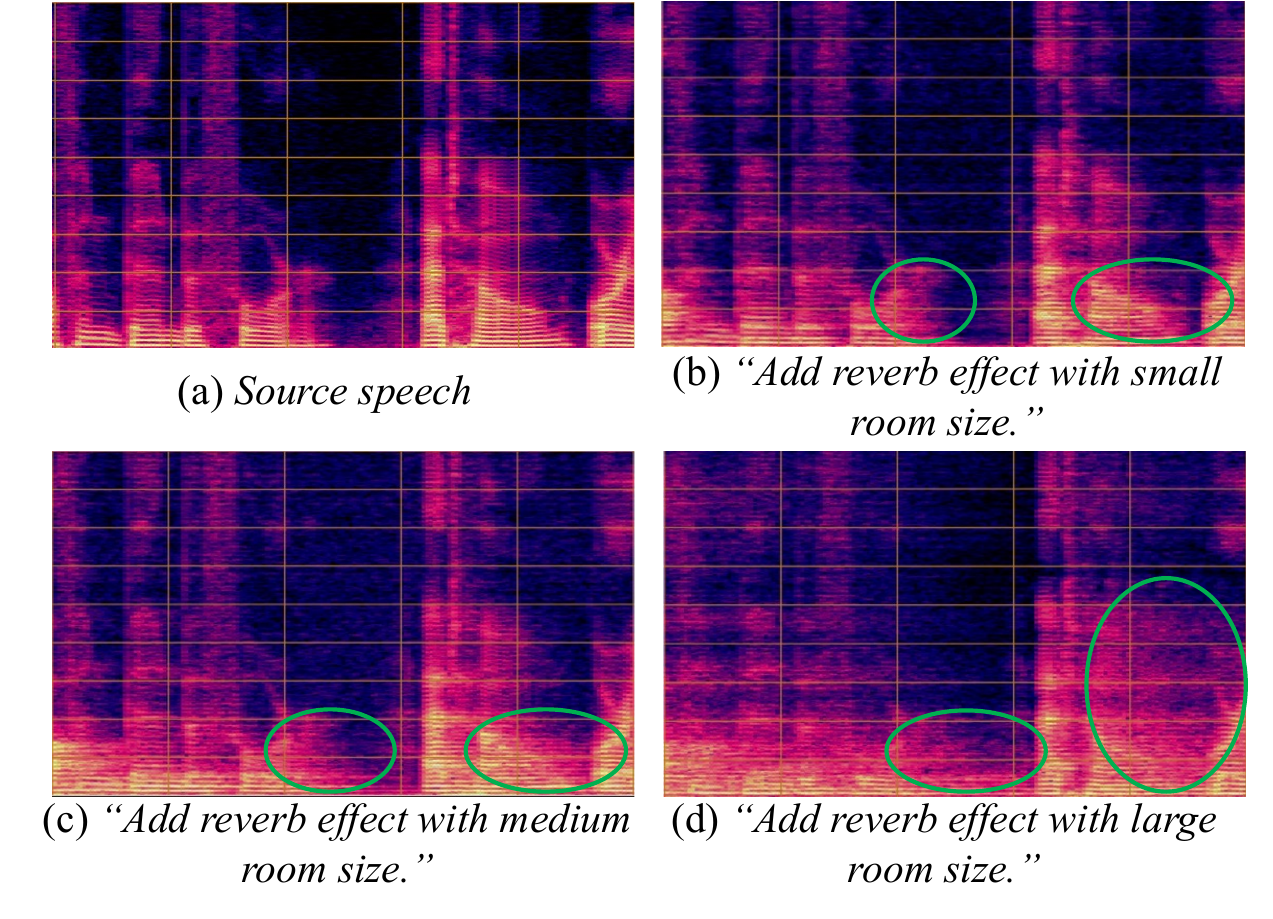}
     \vspace{-4mm}
    \caption{Qualitative demos of edited speech generated by uSee model. Controllable generation is enabled by the textual prompts in the captions to control the room size of RIRs.}
    \vspace{-4mm}
    \label{fig:editing_demo}
\end{figure}

For speech editing, we present the qualitative results of the generated speech in Figure \ref{fig:editing_demo}. We choose to demonstrate that uSee model can achieve controllable generation by showing the spectrograms where various text instructions are applied to the same source speech. The specific text prompts are included in the caption of Figure \ref{fig:editing_demo}, with instructions to introduce reverberation effects with RIRs of different room sizes. From the figure, we can observe that the lengths of reverberant tails in the spectrograms are different according to the text prompts. Examples of reverberant tail differences are highlighted in the green circles. Specifically, Figure \ref{fig:editing_demo}(a) is one example of a source speech, and Figure \ref{fig:editing_demo}(b) prompts the uSee model to add an RIR with small room size to the source speech. Therefore,  a short trail of energy can be observed at the end of the harmonics. Accordingly, if the text prompts are changed to add reverberation effect with medium and large room sizes, as Figure \ref{fig:editing_demo}(c) and (d) show, the length of the reverberant tails are increasingly longer compared to small room size. Since the only difference between the examples lies in the text prompts, we effectively demonstrate the controllable generation of speech editing in our uSee model. In fact, our uSee model can achieve more fine-grained reverberation simulation by conditioning on different RT60s. More demos of the generated speech, e.g., adding background sounds, are available at our project \href{https://muqiaoy.github.io/usee}{website}.

\vspace{-3mm}
\section{Conclusion}
\label{sec:conclusion}
\vspace{-1mm}

This paper presents uSee, a unified speech enhancement and editing framework with a score-based conditional diffusion model. By providing the uSee model with conditions including both acoustic and textual prompts, we show the capability of controllable generation for both speech enhancement and editing tasks. For speech enhancement, the proposed uSee model can yield good performance on speech denoising and dereverberation to generate high-quality speech. In addition, it also demonstrates a fine-grained control of speech editing to add background sound or reverberation effects according to user-defined prompts.




\vfill\pagebreak


\bibliographystyle{IEEEbib}

{\fontsize{9pt}{9pt}\selectfont
\bibliography{refs}
}

\end{document}